\documentclass[a4paper,11pt]{article}
\pdfoutput=1
\usepackage{jcappub}
\usepackage{tikz,xcolor,hyperref}
\usepackage{amsmath, amssymb, amsthm, graphicx, epsfig, fancyhdr,epsfig, slashed,orcidlink}
\usepackage{mathtools}
\usepackage[normalem]{ulem}
\usepackage{tikzsymbols}
\usepackage{natbib}
\usepackage{float}
\usepackage{amsmath}
\usepackage{mathrsfs}
\usepackage{amssymb}
\usepackage{amsfonts}
\usepackage{braket}
\usepackage{booktabs}
\newcommand{\Trh}{T_\text{rh}}
\newcommand{\Tmax}{T_\text{max}}
\newcommand{\mdm}{m_{\rm DM}}
\newcommand{\gs}{g_\star}
\newcommand{\gss}{g_{\star s}}
\newcommand{\ain}{a_{\rm in}}
\newcommand{\Hin}{\mathcal{H}_{\rm in}}
\newcommand{\arh}{a_{\rm rh}}
\newcommand{\Mzp}{m_{Z_{\rm BL}}}
\newcommand{\gbl}{g_{\rm BL}}
\newcommand{\zbl}{Z_{\rm BL}}
\newcommand{\ndm}{n_{\rm DM}}
\newcommand{\Ndm}{N_{\rm DM}}
\newcommand{\Tin}{T_{\rm in}}
\newcommand{\ogw}{\Omega_\text{GW}}
\newcommand{\DNeff}{\Delta N_\text{eff}}
\newcommand{\fmax}{f_\text{max}}

\DeclareUnicodeCharacter{2212}{\textminus}

\begin{document}
\title{An early $Z'$ dominance: dark matter, gravitational waves \& collider probes}
\author[a]{Basabendu Barman\,\orcidlink{https://orcid.org/0000-0003-0374-7655},}
\author[b, d]{Nayan Das,}
\author[c]{Partha Kumar Paul\,\orcidlink{https://orcid.org/0000-0002-9107-5635},}
\author[c]{and Narendra Sahu\,\orcidlink{https://orcid.org/0000-0002-9675-0484}}
\affiliation[a]{\,\,Department of Physics, School of Engineering and Sciences, SRM University AP, Amaravati 522240, India}
\affiliation[b]{\,\,Department of Physics, Indian Institute of Technology Guwahati, Assam- 781039, India}
\affiliation[c]{\,\,Department of Physics, Indian Institute of Technology Hyderabad, Kandi, Telangana-502285, India}
\affiliation[d]{\,\,Centre for High Energy Physics, Indian Institute of Science, C. V. Raman Avenue, Bengaluru 560012, India}
\emailAdd{basabendu.b@srmap.edu.in}
\emailAdd{nayan.das@iitg.ac.in}
\emailAdd{ph22resch11012@iith.ac.in}
\emailAdd{nsahu@phy.iith.ac.in}
\abstract{In the minimal anomaly-free gauged B$-$L extension of the Standard Model (SM), we consider a scenario in which the energy density of the new Abelian gauge boson dominates the Universe prior to big bang nucleosynthesis (BBN), resulting in an early matter-dominated (EMD) era. Taking one of the three right-handed neutrinos to be a decaying dark matter (DM) candidate, we investigate its production during this EMD phase. We show that the region of parameter space consistent with the relic abundance lies within the projected sensitivity of several gravitational wave (GW) experiments, where the GW signal arises from cosmic strings formed during the spontaneous breaking of the gauged B$-$L symmetry. We further constrain the viable parameter space using bounds from collider experiments, beam dump experiments, BBN, and ultra high energy neutrino observations, demonstrating the complementarity of multi-messenger probes in exploring physics beyond the SM as well as non-standard cosmological histories prior to BBN.}
\maketitle
\section{Introduction}
\label{sec:intro}
A wide range of well-motivated beyond the Standard Model (SM) frameworks naturally predict the formation of topological defects, such as cosmic strings and domain walls, during phase transitions in the early Universe~\cite{Vilenkin:1984ib,Vilenkin:2000jqa}. Among these, abelian gauge extensions of the SM~\cite{Davidson:1978pm,Mohapatra:1980qe,Buchmuller:1991ce,Sahu:2005fe,Sahu:2004ir,Montero:2007cd,Langacker:2008yv,Ma:2014qra,Sanchez-Vega:2015qva,Okada:2016gsh,Okada:2016tci,Biswas:2019ygr,Das:2019pua,Ghosh:2021khk,Mahapatra:2025vzu,Ma:2026tyk}, which are theoretically appealing for several reasons, including neutrino mass generation, DM frameworks, and gauge unification, provide a particularly well-studied setting. When such an extended gauge symmetry undergoes spontaneous symmetry breaking, it can give rise to cosmic strings (CS)~\cite{Vilenkin:2000jqa}. Once formed, cosmic strings act as long-lived relics of the early Universe and can efficiently radiate gravitational waves (GWs), producing a stochastic GW background with a characteristic spectral shape~\cite{Vilenkin:1981bx,Turok:1984cn,Caldwell:1996en,Cui:2018rwi,Schmitz:2024gds,Datta:2020bht}. If the symmetry breaking scale is sufficiently high, this GW signal may be detectable by upcoming GW observatories, offering a unique probe of ultra-high-energy physics far beyond the direct reach of collider experiments. In this sense, cosmic strings provide a powerful observational window into new physics beyond the SM at energy scales that would otherwise remain inaccessible. A key prediction of cosmic string networks is an approximately scale-invariant GW spectrum over a broad range of frequencies. Therefore, any departure from this standard behavior can serve as an important indicator of additional new physics or non-standard thermal histories in the early Universe. In particular, the presence of an early matter dominated (EMD) era can significantly modify the expansion history before big bang nucleosynthesis (BBN), imprinting characteristic spectral breaks or distortions in the GW spectrum generated by cosmic strings~\cite{Cui:2017ufi,Cui:2018rwi,Gouttenoire:2019kij}. Such features make the cosmic string GW background not only a probe of new physics beyond the SM, but also sensitive to the pre-BBN cosmological evolution.

Motivated by these considerations, we study a scenario in which the early Universe experiences an epoch of early matter domination driven by a heavy Abelian gauge boson, $\zbl$, arising from a gauged $U(1)_{B-L}$ extension of the SM. In this framework, the cancellation of gauge anomalies naturally requires the introduction of three singlet right-handed neutrinos (RHNs)~\cite{Davidson:1978pm,Mohapatra:1980qe,Buchmuller:1991ce,Sahu:2005fe,Sahu:2004ir,Montero:2007cd,Langacker:2008yv,Das:2015nwk}. We identify one of these RHNs as a viable dark matter (DM) candidate, while the other two generate the observed light neutrino masses through the seesaw mechanism \cite{Sanchez-Vega:2015qva,Sahu:2005fe,Paul:2024iie}. Within this setup, we investigate DM production during the $\zbl$-dominated epoch. The presence of this non-standard phase of cosmological evolution significantly modifies the DM yield and, as a result, changes the allowed parameter space compared to the standard thermal cosmology. Compared to the standard cosmology, the early matter-dominated era typically requires a larger coupling between the DM and the visible sector to reproduce the observed relic abundance. The amount of enhancement in the interaction strength depends on the mediator mass ($\Mzp$) and the DM mass ($\mdm$), leading to improved detection prospects for the model at several experimental frontiers. We perform a systematic analysis of the resulting parameter space in the gauge coupling ($\gbl$)--mass ($\Mzp$) plane, taking into account constraints from collider searches and low-energy experiments. It is well known that the $[\gbl-\Mzp]$ parameter space of the B$-$L extension is already strongly constrained by a wide range of experimental and observational data. We show that the $\zbl$-driven early matter-dominated phase provides a cosmological framework in which the same phenomenological parameter space can also be used to constrain the duration of early matter domination. This allows current observations to probe possible deviations from the standard cosmological history prior to BBN. 

The paper is structured as follows. In Sec.~\ref{sec:framework}, we introduce the underlying particle physics model of the framework. Sec.~\ref{sec:EMD} is devoted to a detailed discussion of the $\zbl$-induced EMD epoch, along with the associated DM production mechanism. In Sec.~\ref{sec:GW}, we analyze how this EMD phase modifies the cosmic string-generated GW spectrum. The resulting viable parameter space, consistent with all relevant constraints and phenomenological requirements, is presented in Sec.~\ref{sec:result}. Finally, we summarize our main findings and conclude in Sec.~\ref{sec:concl}.
\section{Particle content \& interactions}
\label{sec:framework}
We consider a minimal U(1)$_{B-L}$ gauge extension of the SM, whose particle content is listed in Tab.~\ref{table1}. The global $B-L$ symmetry of the SM is promoted to a local gauge symmetry. To ensure anomaly cancellation, three right-handed neutrinos are introduced. A complex scalar $\Phi$ with $B-L$ charge $+2$ is added to spontaneously break the U(1)$_{B-L}$ symmetry. The Yukawa interactions relevant for neutrino masses are given by
\begin{align}
\mathcal{L}_{\rm Yukawa} \supset 
& - Y_\nu\,\overline{\ell_{L}}\,H\, N_R-Y_N\,\Phi\,\overline{N_R^{C}}\,N_R + {\rm h.c.}\,,
\label{eq:Lyuk}
\end{align}
The first term corresponds to the Dirac Yukawa interactions, while the remaining terms generate Majorana masses after U(1)$_{B-L}$ breaking. Without loss of generality, the Majorana Yukawa matrix is taken to be diagonal. Note that here we have suppressed the generational index. 
\begin{table}[htb!]
\centering
\renewcommand{\arraystretch}{1.4}
\setlength{\tabcolsep}{12pt}
\begin{tabular}{c|c c c c}
\toprule
Fields& SU(3)$_c$  & SU(2)$_L$ & U(1)$_Y$ & U(1)$_{B-L}$ \\
\hline\hline
$q^{i}_{L}$ & {\bf 3 } & {\bf 2} & $ 1/6$ & $1/3$ \\
$u^{i}_{R}$ & {\bf 3 } & {\bf 1} & $ 2/3$ & $1/3$ \\
$d^{i}_{R}$ & {\bf 3 } & {\bf 1} & $-1/3$ & $1/3$ \\
\hline
$\ell^{i}_{L}$ & {\bf 1 } & {\bf 2} & $-\frac{1}{2}$ & $-1$ \\
$e^{i}_{R}$ & {\bf 1 } & {\bf 1} & $-1$ & $-1$ \\
\hline
$H$ & {\bf 1 } & {\bf 2} & $\frac{1}{2}$ & $0$ \\ 
\hline
$N^{j}_{R}$ & {\bf 1 } & {\bf 1} & $0$ & $-1$ \\
$\Phi$ & {\bf 1 } & {\bf 1} & $0$ & $+2$ \\
\bottomrule
\end{tabular}
\caption{Particle content of the minimal U(1)$_{B-L}$ extension of the SM.}
\label{table1}
\end{table}
The most general renormalizable scalar potential involving $H$ and $\Phi$ is
\begin{align}
V &= -\mu_H^2 (H^{\dagger}H)- \mu_\Phi^2 (\Phi^{\dagger} \Phi)+ \lambda_H (H^{\dagger}H)^2+ \lambda_{\Phi} (\Phi^{\dagger} \Phi)^2 
+ \lambda_{H\Phi} (H^{\dagger}H)(\Phi^{\dagger} \Phi)\,.
\label{eq:pot}
\end{align}
The parameters are chosen such that both scalars develop non-vanishing vacuum expectation values (VEVs),
\begin{align}
\langle H \rangle &=\frac{1}{\sqrt{2}}  
\begin{pmatrix}  
0
\\[10pt]
h+v_h 
\end{pmatrix},
\qquad
\langle \Phi \rangle = \frac{\phi+v_\phi}{\sqrt{2}}\,.
\end{align}
After U(1)$_{B-L}$ breaking, the Majorana masses of the right-handed neutrinos (RHNs) and the mass of the new gauge boson $\zbl$ are given by
\begin{align}\label{eq:mass}
& m_N=\sqrt{2}\,{Y_N}\,v_\phi, & 
\Mzp = 2\,\gbl\,v_\phi\,,
\end{align}
where $\gbl$ denotes the U(1)$_{B-L}$ gauge coupling. 

Once both $B-L$ and electroweak symmetries are spontaneously broken, the two scalar fields mix. The mixing angle is given by
\begin{align}
\theta=\frac{1}{2}\,\arctan\left[\frac{v_h\,v_\phi\,\lambda_{H\Phi}}
{2\,v_\phi^2\,\lambda_\Phi-2\,v_h^2\,\lambda_H}\right]\,.
\end{align}
In the limit $v_\phi \gg v_h$, this expression simplifies to
\begin{align}
\theta \simeq \frac{\lambda_{H\Phi}\,v_h}{2\,\lambda_\Phi\,v_\phi}\,.
\end{align}
Thus, for $v_\phi \gg v_h$ and moderately small $\lambda_{H\Phi}$, the scalar mixing becomes naturally suppressed. Using $m_\Phi^2 \simeq 2\lambda_\Phi v_\phi^2=2\mu_\Phi^2$, the mixing angle can equivalently be written as,
\begin{align}
\theta \simeq \frac{\lambda_{H\Phi}\,v_h\,v_\phi}{m_\Phi^2}\,.
\end{align}
We work in the limit of vanishing portal coupling $\lambda_{H\Phi}\to0$ such that $\theta\to0$, and hence the Higgs-mediated interactions can be assumed to be absent. Among the three RHNs, we consider $N_1$ to be a good DM candidate with mass around the GeV--PeV scale, while $N_{2,3}$ can take part in light neutrino mass generation via type-I seesaw and leptogenesis\cite{Buchmuller:2004nz}. We primarily focus on $N_1$ production from the SM bath via $\zbl$ mediation\footnote{This can be realized by considering $m_\Phi\gg\Tmax$, where $\Tmax$ is the maximum temperature of the Universe during reheating. As a result, $\Phi$ is never thermally populated during reheating. Consequently, scalar-mediated contributions to DM production are strongly suppressed.}, during an epoch of early matter domination, the details of which shall be discussed in the next section. Since $N_1$ plays the role of a DM candidate, here onward we will denote $N_1$ mass as $\mdm$. We do not impose any additional stabilizing symmetry to forbid DM decay. Instead, the DM is considered to be long-lived by choosing a sufficiently small Yukawa coupling $Y_{\nu_1}$. A simple estimate gives
\begin{align}
\tau_{N_1}\simeq\left(\frac{Y_{\nu_1}^2\,\mdm}{8\pi}\right)^{-1}\simeq 4.1\times 10^{17}\,\text{s}\,\left(\frac{2\times 10^{-22}}{Y_{\nu_1}}\right)^2\,\left(\frac{1\,\text{TeV}}{\mdm}\right)\,,
\end{align}
which shows that an extremely suppressed Yukawa coupling is needed. In our framework, since the lightest active neutrino is massless, this coupling can always be adjusted to ensure a sufficiently long-lived $N_1$. 
\section{Cosmology during $\zbl$  domination}
\label{sec:EMD}
In this section, we examine the cosmological implications of an EMD epoch driven by the energy density of the $\zbl$ field. We do not attempt to construct the underlying inflationary scenario that produces the initial $\zbl$-dominated state\footnote{Inflation driven by massive vector fields and their subsequent reheating have been studied in the literature; see, e.g., Refs.~\cite{Ford:1989me,Golovnev:2008cf,Clery:2023mjo}.}. We assume that the post-inflationary Universe is initially dominated by the energy density of the long-lived $Z_{\rm BL}$ field, giving rise to an EMD epoch. The SM radiation bath is generated solely through the {\it perturbative} decay of $Z_{\rm BL}$, thereby reheating the Universe at the end of $\zbl$-domination. The process of reheating completes sufficiently early to preserve the successful predictions of standard BBN, which takes place when the thermal bath temperature $\simeq 4~\mathrm{MeV}$~\cite{Sarkar:1995dd, Kawasaki:2000en, Hannestad:2004px, DeBernardis:2008zz, deSalas:2015glj,Hasegawa:2019jsa}.
\subsection{Evolution of energy densities}
The evolution of the SM radiation and the energy density of the $\zbl$ field can be consistently followed using a coupled set of Boltzmann equations (BEQs), 
\begin{align}
& \frac{d\rho_{\zbl}}{dt} + 3\,\mathcal{H}\,\rho_{\zbl}=-\Gamma_{\zbl}\,\rho_{\zbl}\,,
\nonumber\\&
\frac{d\rho_R}{dt} + 4\mathcal{H}\rho_R = +\Gamma_{\zbl}\,\rho_{\zbl}\,.
\label{eq:beq}
\end{align}
where $\Gamma_{\zbl}$ is the total decay width of $\zbl$. The photon temperature can be extracted from the SM energy density via $\rho_R=(\pi^2/30)\,\gs(T)\,T^4$, where $\gs(T)$ is the number of relativistic degrees of freedom (DoF) contributing to the total radiation energy density. The Hubble expansion rate $\mathcal{H}$ receives contributions from all energy components, namely both the SM radiation ($\rho_R$) and the $\zbl$ field ($\rho_{\zbl}$), such that
\begin{align}
\mathcal{H} = \sqrt{\frac{\rho_{\zbl}+\rho_R}{3\,M_P^2}}\,,
\end{align}
where, $M_P\simeq2.43\times10^{18}$ GeV is the reduced Planck mass. Considering instantaneous thermalization of the SM particles, the reheating temperature $\Trh$ can be defined via the relation,
\begin{align}
\Trh^2=\frac{3}{\pi}\,\sqrt{\frac{10}{\gs(\Trh)}}\,M_P\,\Gamma_{\zbl}\,.
\label{eq:GammaZp}
\end{align}
During EMD, the Universe is dominated by the $\zbl$ field. Consequently, the term associated with the expansion   $3\,\mathcal{H}\,\Gamma_{\zbl}$ typically dominates over the interaction term $\Gamma_{\zbl}\,\rho_{\zbl}$. The right hand side of the first BEQ can thus be neglected and one obtains the evolution of $\zbl$ energy density as,
\begin{align}
\rho_{\zbl}(a)
= \rho_{\zbl}^{\rm in}
\left(\frac{a_{\rm in}}{a}\right)^3
= 3M_P^2 \mathcal{H}_{\rm in}^2
\left(\frac{a_{\rm in}}{a}\right)^3\,,
\label{eq:rhoZp}
\end{align}
where $\mathcal{H}_{\rm in} \equiv \mathcal{H}(a_{\rm in})\simeq\sqrt{\rho_{\zbl}^{\rm in}/(3M_P^2)}$ and $\ain$ denotes an arbitrary initial scale factor. Substituting Eq.~\eqref{eq:rhoZp} in the second BEQ, we obtain the evolution of radiation energy density as,
\begin{align}
\rho_R(a)
= \frac{6}{5}\,
M_P^2\,\mathcal{H}_{\rm in}\,\Gamma_{\zbl}\,
\left(\frac{a_{\rm in}}{a}\right)^{3/2}\,\left[1-
\left(\frac{a_{\rm in}}{a}\right)^{5/2}
\right],
\label{eq:rhoR}
\end{align}
where we have imposed the initial condition $\rho_R(a_{\rm in})\to0$. The thermal bath attains a maximum temperature $\Tmax$ when only a small fraction of the energy stored in the massive mediator $\zbl$ has decayed into radiation. The maximum temperature occurs at the scale factor
\begin{align}
a_{\max}
= a_{\rm in}
\left(\frac{8}{3}\right)^{\!2/5}\,,
\label{eq:amax}
\end{align}
with corresponding temperature
\begin{align}\label{eq:Tmax}
&\Tmax^4=\left(\frac{3}{8}\right)^{3/5}\,\frac{45\,M_P^2}{2\,\pi^2\,\gs(\Tmax)}\,\Hin\,\Gamma_{\zbl}\,.  
\end{align}
During the epoch dominated by the decaying $\zbl$ field, the temperature of the thermal bath scales as,
\begin{align}
& T(a) \propto a^{-3/8}\,, & a_{\max}\leq a < \arh\,,
\end{align}
where, 
\begin{align}
& \arh=\ain\,\left(\frac{5}{2}\,\frac{\Hin}{\Gamma_{\zbl}}\right)^{2/3}\,,
\end{align}
denotes the epoch at which the equality $\rho_{\zbl}(\arh)=\rho_R(\arh)$ holds. Once radiation domination is established, the temperature follows the standard scaling $T(a)\propto a^{-1}$. The dynamics of the EMD epoch are therefore controlled by the following set of free parameters: 
\begin{align}\label{eq:freepar}
\Big\{\ain,\,\mathcal{H}_{\rm in},\,\Gamma_{\zbl}\,(\gbl,\,\Mzp)\Big\}\,.    
\end{align}
Once again, from Eq.~\eqref{eq:freepar} it is evident that the EMD epoch is completely decided by the underlying microphysics.
\subsection{Dark matter during EMD}
During the epoch of EMD, the DM can be produced non-thermally~\cite{McDonald:2001vt,Hall:2009bx,Bernal:2017kxu} via: (i) on-shell decay of $\zbl$ and (ii) $\zbl$-mediated annihilation of the SM states. Detailed expressions for all relevant decay rates and annihilation cross-sections are provided in Appendix~\ref{sec:decay-cross-section}. Now, the DM production rate from $\zbl$ decay is given by the production rate densities as,
\begin{align}
\gamma_{1\to2}=\frac{g_{\zbl}}{2\,\pi^2}\,\Mzp^2\,\Gamma_{\zbl\to \text{DM}\,\text{DM}}\,T\,K_1\left(\frac{\Mzp}{T}\right)\,,
\end{align}
where $g_{\zbl}$ is the DoF of $\zbl$ and $K_1[...]$ is the first order modified Bessel function. On the other hand,  production rate from 2-to-2 scattering off of the bath particles reads,
\begin{align}
& \gamma_{2\to2}=\frac{T}{32\pi^4}\,g_a g_b
\nonumber\\&
\times\int_{\text{max}\left[\left(m_a+m_b\right)^2,4\mdm^2\right]}^\infty ds\frac{\biggl[\bigl(s-m_a^2-m_b^2\bigr)^2-4\,m_a^2 m_b^2\biggr]}{\sqrt{s}}\,\sigma\left(s\right)_{a,b\to\text{DM},\,\text{DM}}\,K_1\left(\frac{\sqrt{s}}{T}\right)\label{eq:gam-ann}\,. 
\end{align}
Here, the sub-indices $a,b$ correspond to the SM states, $g_{a,b}$ are the corresponding DoF, for Dirac fermions $g_{a,b}=2\,N_c$, where $N_c=3(1)$ for quarks(leptons).
\begin{figure}
    \centering    \includegraphics[scale=0.15]{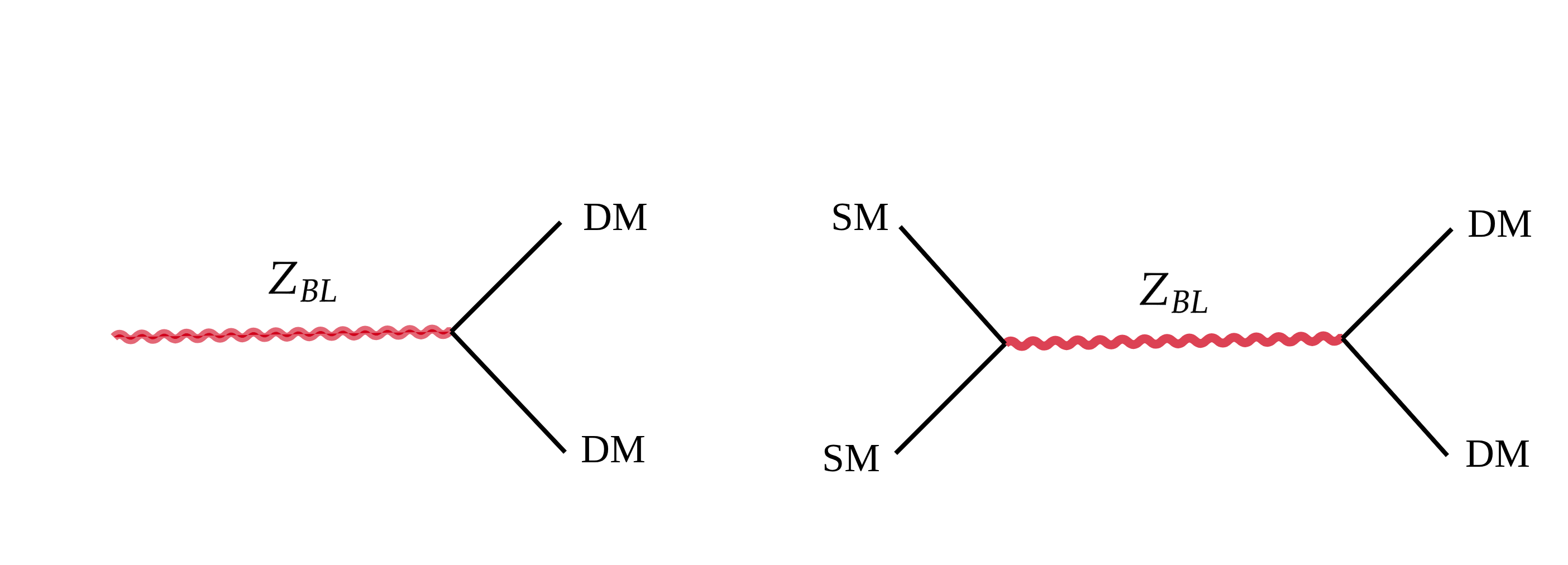}
    \caption{Feynman diagrams corresponding to DM production from $\zbl$ decay (left) and $\zbl$ mediated scattering of the SM particles in the thermal bath (right).}
    \label{fig:feyn}
\end{figure}

To analytically understand the behavior of the reaction density via 2-to-2 $\zbl$-mediated scattering in the $s$-channel, we work in the massless limit of the initial and final state particles. We emphasize that in the numerical analysis, we have taken into account all possible DM production channels. In this limit, the total DM production cross-section is given by,
\begin{align}
&\sigma(s)\simeq \frac{7\,\gbl^4}{27\,\pi}\,\frac{s}{\left(s-\Mzp^2\right)^2+\Gamma_{\zbl}^2\,\Mzp^2}\,.
\end{align}
As a result the reaction density in Eq.~\eqref{eq:gam-ann} simplifies to,
\begin{align}
&\gamma_{2\to2}\simeq g_a\,g_b\frac{T}{32\,\pi^4}\,\int_{4\,\mdm^2}^\infty ds\,s^{3/2}\,\sigma(s)\,K_1\left(\frac{\sqrt{s}}{T}\right)\,. 
\end{align}
Near the resonance, the narrow width approximation $\frac{1}{(s-\Mzp^2)^2+\Gamma_{\zbl}^2\,\Mzp^2}\to\frac{\pi}{\Mzp\,\Gamma_{\zbl}}\,\delta(s-\Mzp^2)$ is applicable. Away from the resonance, the propagator becomes $1/\left(s-\Mzp^2\right)^2$, ignoring the mediator decay rate. With the approximations and assumptions furnished, we obtain an approximate analytical expression for the reaction density of 2-to-2 scattering as,
\begin{align}\label{eq:gam-ann2}
& \gamma_{2\to2}\simeq\gbl^4
\begin{dcases}
\frac{32}{\pi^5}\,\frac{T^8}{\Mzp^4}\,, & T\ll\Mzp/2\,,
\\[10pt]
\frac{1}{24\,\pi^4}\,\frac{\Mzp^4\,T}{\Gamma_{\zbl}}\times K_1\left(\Mzp/T\right)\,, & T\simeq\Mzp/2\,,
\\[10pt]
\frac{T^4}{6\,\pi^5}\,, & T\gg\Mzp/2\,,
\end{dcases}
\end{align}
which shows a strong temperature-dependence typically in the regime $T\ll\Mzp/2$. 
\begin{figure}[htb!]
\centering
\includegraphics[scale=0.36]{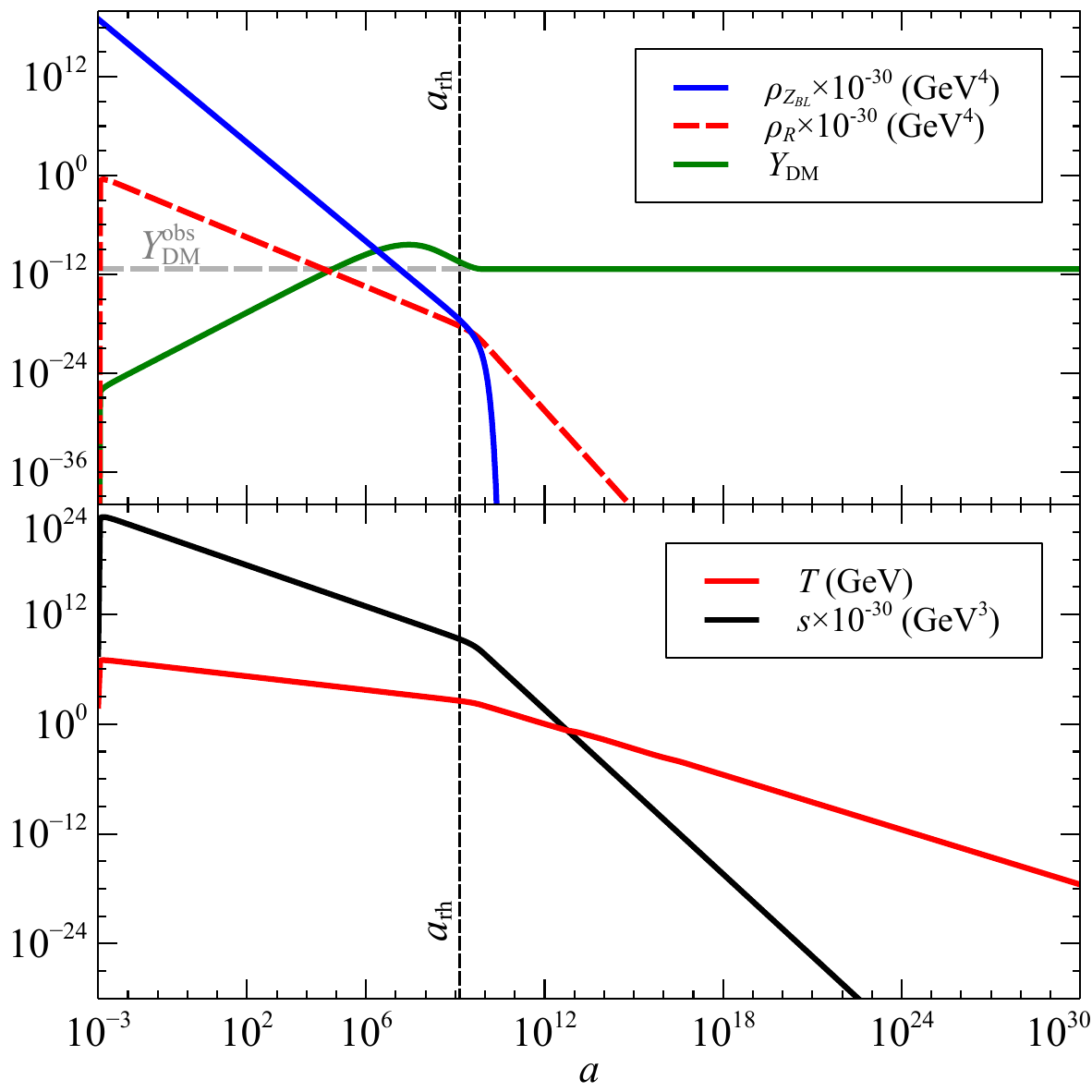}~
\includegraphics[scale=0.36]{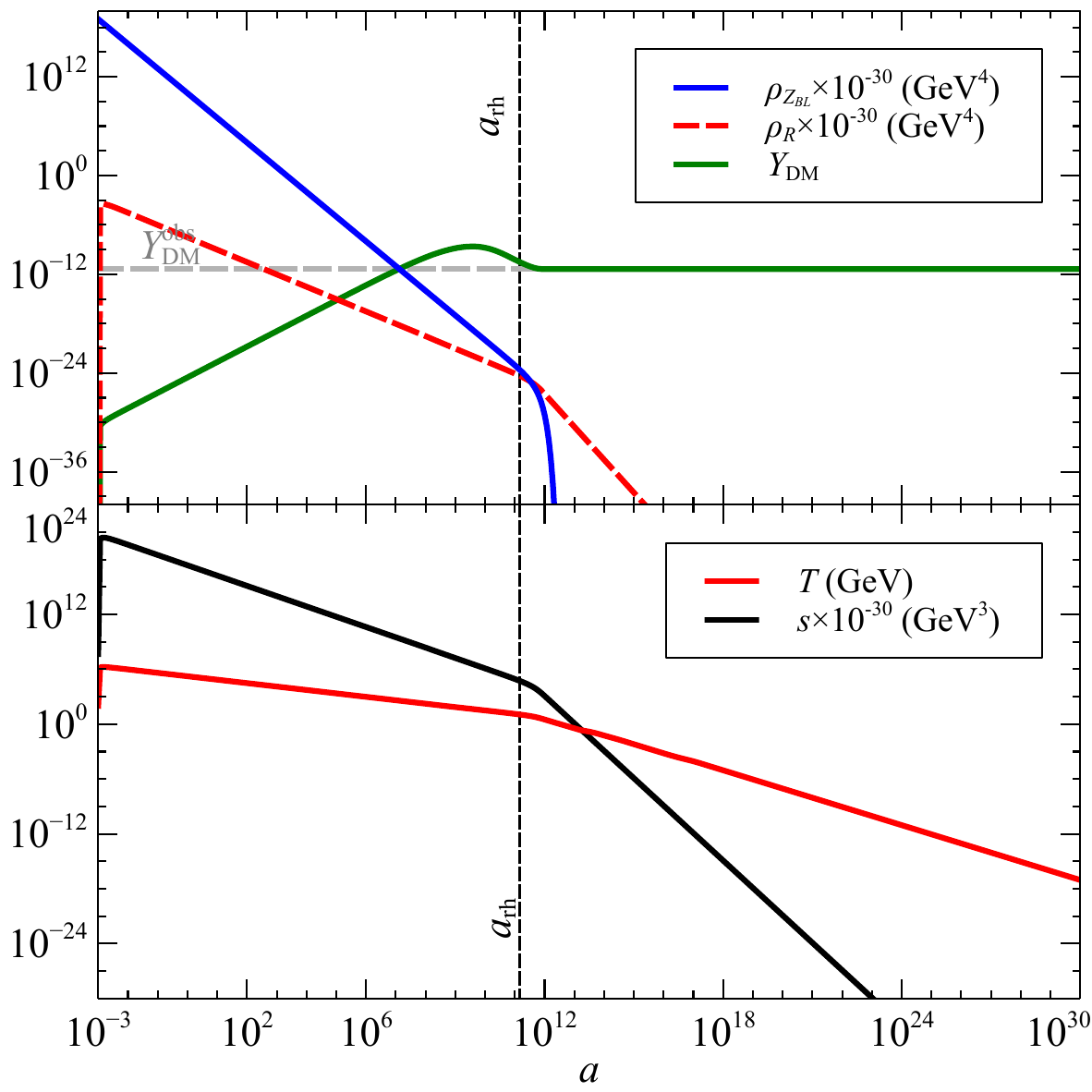}
\caption{Left: Evolution of $\zbl$ (blue), radiation (red dashed), DM (green) energy densities along with bath temperature (red solid) and entropy density (black solid), as a function of the scale factor. In all cases we have considered: $\Mzp=10^{4}~{\rm GeV},\,\gbl\simeq 9\times10^{-9},\,\Hin=7.5\times10^5~\text{GeV},\mdm=100\,\text{GeV}$. Right: Same as top panel, but with $\Mzp=300~{\rm GeV},\,\gbl\simeq1.6\times10^{-9},\,\mdm=100\,\text{GeV}$.}
    \label{fig:evol}
\end{figure}

To track the evolution of DM number during the EMD era, we solve the following BEQ taking into all possible DM production channels,
\begin{align}\label{eq:beq}
&\frac{d\Ndm}{da}=\frac{a^2}{\mathcal{H}}\,\gamma_{\rm DM}\,, 
\end{align}
where $\Ndm\equiv\ndm\,a^3$ is the comoving DM number and $\gamma_{\rm DM}\equiv(\gamma_{1\to2}+\gamma_{2\to2})$ corresponds to the total DM production rate density from $\zbl$ decay as well as $\zbl$-mediated scattering. Even though the SM entropy density is not conserved when $\zbl$ is decaying, the DM yield can be defined as,
\begin{align}
Y(T)\equiv\frac{\Ndm(T)}{\mathfrak{s}(T)\,a^3}\,,  
\end{align}
where $\mathfrak{s}(T)=(2\pi^2/45)\,\gss(T)\,T^3$ is the entropy energy density, with $\gss(T)$ being the number of releativistic DoFs contributing to the entropy density. Reproducing the observed DM relic density requires $Y_0\, \mdm = \Omega h^2 \, \frac{1}{s_0}\,\frac{\rho_c}{h^2} \simeq 4.3 \times 10^{-10}\,\text{GeV}$, where $Y_0=Y(T_0)$ is the yield at present temperature $T_0$. Here we use the critical energy density $\rho_c \simeq 1.05 \times 10^{-5}\, h^2$~GeV/cm$^3$, the present entropy density $\mathfrak{s}_0\simeq 2.69 \times 10^3$~cm$^{-3}$~\cite{ParticleDataGroup:2022pth}, and the DM relic abundance $\Omega h^2 \simeq 0.12$, with $h\simeq H_0/100\,\left({\rm km/s/Mpc}\right)$ the reduced Hubble parameter, where $H_0\simeq 67.4 \pm 0.5 \text{km/s/Mpc}$ is the present Hubble rate~\cite{Planck:2018vyg}. For DM production entirely from on-shell $\zbl$ decay (if $\Mzp>2\,\mdm$) Eq.~\eqref{eq:beq} admits an approximate analytical solution that reads,
\begin{align}\label{eq:Ndm12}
& \Ndm(\arh)^{1\to2}\simeq \frac{\ain^2\,\gbl^2}{3\times 6^{1/4}\,\pi^3}\,\frac{\Tmax^{28/3}}{\Mzp^{16/3}\,\Hin}\,\left[\Gamma \left(\frac{53}{6},x_{\rm in}\right)-\Gamma \left(\frac{53}{6},x_{\rm rh}\right)\right]\,, 
\end{align}
with an asymptotic yield
\begin{align}
&  Y_0^{1\to2}\equiv\frac{\Ndm(\arh)^{1\to2}}{\arh^3\,s(\arh)}\simeq\frac{5}{2}\,\left(\frac{27}{2}\right)^{1/4}\,\frac{\gbl^2\,\Tmax^{28/3}}{\ain\,\Hin\,\Mzp^{16/3}\,\Trh^3}\,\left(\frac{\ain}{\arh}\right)^3\,\frac{\left[\Gamma \left(\frac{53}{6},x_{\rm in}\right)-\Gamma \left(\frac{53}{6},x_{\rm rh}\right)\right]}{\gss\,\pi^5}\,,    
\end{align}
where $x\equiv \left(a/a_{\rm max}\right)^{3/8}\,\Mzp/\Tmax$, $\Gamma[a,\,z]$ is the incomplete gamma-function and $\Tmax$ is given by Eq.~\eqref{eq:Tmax}. On the other hand, for $\Mzp<2\,\mdm$, the DM can only be produced via $\zbl$-mediated process. In that case, using Eq.~\eqref{eq:gam-ann2} we obtain,
\begin{align}\label{eq:Ndm22}
& \Ndm(\arh)^{2\to2}\simeq\frac{\ain^3\,\gbl^4}{\Hin}
\begin{dcases}
\frac{8960}{9\,\pi^5}\,\frac{\Tmax^8}{\Mzp^4}\,\left(\frac{\arh}{\ain}\right)^{3/2}\,, & T\ll\Mzp/2\,,
\\[10pt]
\frac{140}{27\,\pi^4}\,\left(\frac{8}{3}\right)^{1/4}\,\frac{\Tmax^{28/3}}{\Mzp^{13/3}\,\Gamma_{\zbl}\,\ain}\,\mathcal{F}\,, & T\simeq\Mzp/2\,,
\\[10pt]
\frac{70}{9\,\pi^5}\,\Tmax^4\,\left(\frac{\arh}{\ain}\right)^3\,, & T\gg\Mzp/2\,,
\end{dcases}
\end{align}
where $\mathcal{F}\equiv \left[\Gamma \left(\frac{53}{6},x_{\rm in}\right)-\Gamma \left(\frac{53}{6},x_{\rm rh}\right)\right]$. One then arrives at the final yield that reads,
\begin{align}\label{eq:Ydm}
& Y_0^{2\to2}\equiv\frac{\Ndm(\arh)^{2\to2}}{\arh^3\,s(\arh)}=\frac{45\,\gbl^4}{2\pi^2\,\gss(\Trh)}
\begin{dcases}
\left(\frac{5}{2}\right)^{1/3}\,\frac{8960}{9\pi^5}\,\left(\frac{\Gamma_{\zbl}}{\Hin}\right)^{1/6}\,\frac{\Tmax^8}{\Hin\,\Mzp^4\,\Trh^3}\,, & T\ll\Mzp/2\,,
\\[10pt]
\frac{875}{27\,\pi^4\,\ain}\,\left(\frac{8}{3}\right)^{1/4}\,\frac{\Tmax^{28/3}}{\Mzp^{13/3}\,\Hin^2\,\Trh^3}\,\mathcal{F}\,, & T\simeq\Mzp/2\,,
\\[10pt]
\frac{70}{9\pi^5}\,\frac{\Tmax^4}{\Hin\,\Trh^3}\,, & T\gg\Mzp/2\,.
\end{dcases}
\end{align}
Note that, in this case, the yield is independent of the mediator mass in the limit $T\gg\Mzp/2$, while it is suppressed by $\Mzp$ for $T\ll\Mzp/2$.

The evolution of the energy densities of $\zbl$, radiation, and DM is shown in the top left and top right panel of Fig.~\ref{fig:evol}. For the top left panel, the masses are chosen in such a way that DM production dominantly takes place from $\zbl$ decay. During the EMD era, ($\ain < a < \arh$) the energy density of $\zbl$ clearly dominates over both radiation and DM, as expected. The decay of $\zbl$ reheats the Universe to a temperature $\Trh\sim 357~\text{GeV}$, as illustrated in the bottom left panel. While $\zbl$ dominates, the SM bath temperature follows the characteristic scaling $T \propto a^{-3/8}$. For $a > \arh$, once the EMD is over and radiation domination sets in, the temperature scales as $T \propto a^{-1}$, as expected for freely redshifting radiation. Since $m_{\rm DM} < \Mzp/2$, DM production in this scenario is dominated by the decay of $\zbl$, rather than by scattering channels. Contrary to the standard radiation-domination, here we see, after freeze-in, the comoving DM number density is diluted due to entropy injection until $T\sim\Trh$. Once the decay of $\zbl$ effectively ceases and reheating completes, entropy production shuts off and the DM abundance saturates to its final observed value. Similar dynamics is also visible in the top right panel, where we consider $m_{\zbl}=300$ GeV and $\mdm=100{~\rm GeV}$.    

It is essential for freeze-in that the DM production rate remains below the Hubble rate during production. This can be ensured by demanding,
\begin{eqnarray}
\mathbb{R}\equiv\frac{\gamma_{\rm DM}}{n_{\rm DM}^{\rm eq}\,\mathcal{H}}<1\,.
\end{eqnarray}
Note that, here we use equilibrium number density for the DM to obtain a more conservative constraint on the relevant parameters. This provides an upper bound on $\gbl$,
\begin{align}\label{eq:gbl}
&\gbl\lesssim
\begin{dcases}
\left(\frac{3\,\pi^3}{35}\right)^{1/4}\,\left(\frac{\Hin}{T^3}\right)^{1/4}\,\left(\frac{\mdm
\,\Mzp^2}{16\,\Tin^2}\right)^{1/2}\,K_2\left(\frac{\mdm}{T}\right)^{1/4}\,, & T \ll \Mzp/2\,,
\\[10pt]
\sqrt{3\,\pi}\,\left(\frac{\Hin\,\Gamma_{\zbl}}{35}\right)^{1/4}\,\left(\frac{\mdm\,T^2}{\Mzp^2\,\Tin^2}\right)^{1/2}\,\left[\frac{K_2\left(\mdm/T\right)}{K_1\left(\Mzp/T\right)}\right]^{1/4}\,, & T \simeq\Mzp/2\,,
\\[10pt]
\sqrt{\frac{3}{2}}\,\left(\frac{\pi^3}{35}\right)^{1/4}\,\left(\frac{\Hin\,\mdm^2\,T}{\Tin^4}\right)^{1/4}\,K_2\left(\frac{\mdm}{T}\right)^{1/4}\,, & T \gg \Mzp/2\,.
\end{dcases}
\end{align}

Based on the above analysis, we identify six benchmark points (BPs) in Tab.~\ref{tab:BP} that reproduce the observed relic abundance. For each BP, we also list the $U(1)$ breaking scale $v_\phi$ and the temperature $T_\Delta$ marking the end of the EMD era. These parameters are essential for discussing the gravitational wave signals from cosmic strings during the $Z'$-dominated epoch, as presented in the subsequent section.
\begin{table}[htb!]
\centering
\renewcommand{\arraystretch}{1.2}
\setlength{\tabcolsep}{10pt}
\begin{tabular}{c|c c c c c c}
\toprule
\textbf{BPs} & $\boldsymbol{g_{\rm BL}}$ & $\boldsymbol{m_{Z_{\rm BL}}}$ [GeV] & $\boldsymbol{\Gamma_{Z_{\rm BL}}^{-1}}$ [s] & $\boldsymbol{v_\phi}$ [GeV] & $\boldsymbol{T_{\Delta}}$ [GeV] \\
\hline\hline
BP0 & $9.03\times10^{-11}$ & $0.71$       & $6.8\times10^{-4}$ & $3.94\times10^9$  & $2.89\times10^{-2}$ \\
BP1 & $7.88\times10^{-6}$ & $10^{-3}$       & $1.33\times10^{-10}$ & $63.45$  & $39.23$ \\
BP2 & $3.17\times10^{-6}$ & $10$       & $3.26\times10^{-14}$ & $1.58\times10^{6}$  & $2.49\times10^{3}$ \\
BP3 & $8.92\times10^{-9}$ & $10^{4}$   & $3.67\times10^{-12}$ & $5.61\times10^{11}$ & $2.29\times10^{2}$ \\
BP4 & $8.92\times10^{-8}$ & $10^{6}$   & $3.67\times10^{-16}$ & $5.61\times10^{12}$ & $2.29\times10^{4}$ \\
BP5 & $8.92\times10^{-7}$ & $10^{8}$   & $3.67\times10^{-20}$ & $5.61\times10^{13}$ & $2.29\times10^{6}$ \\
BP6 & $8.92\times10^{-6}$ & $10^{10}$  & $3.67\times10^{-24}$ & $5.61\times10^{14}$ & $2.29\times10^{8}$ \\
\bottomrule
\end{tabular}
\caption{Benchmark points yielding the correct DM abundance. For BP1-BP6, DM mass is fixed at $\mdm = 100\,\mathrm{GeV}$, whereas for BP0, $m_{\rm DM}=200$ MeV. For each case, $T_{\rm \Delta}$ denotes the temperature at which radiation era begins, and $\Gamma_{Z_{\rm BL}}^{-1}$ is the $\zbl$ lifetime.}
\label{tab:BP}
\end{table}
\section{Modified gravitational wave signatures from cosmic strings}
\label{sec:GW}
We now discuss the stochastic gravitational wave (GW) spectrum generated by a cosmic string (CS) network. Cosmic strings are one-dimensional topological defects that can form during the spontaneous breaking of a continuous symmetry when the vacuum manifold is not simply connected~\cite{Vilenkin:2000jqa}. At the microscopic level, a cosmic string network is characterized by its correlation length, $L \equiv \sqrt{\mu/\rho_{\infty}}$, where $\mu$ is the string tension (energy per unit length) and $\rho_{\infty}$ is the energy density of long strings. The string tension is typically set by the symmetry-breaking scale, $\Lambda_{\rm CS}$, with $\mu \sim \Lambda_{\rm CS}^2$. It is often expressed in dimensionless form as $G\mu$, where $G$ is Newton’s gravitational constant. Cosmic strings form randomly and initially create a network. Soon after formation, their motion is damped through interactions with the thermal plasma~\cite{Vilenkin:2000jqa}. Once this friction-dominated phase ends, the network evolves under two competing effects: (i) cosmic expansion stretches the correlation length, and (ii) long strings continuously produce loops. These processes drive the network into the well-known scaling regime~\cite{Vilenkin:2000jqa}, where the correlation length grows with cosmic time as $L \sim t$. Consequently, for constant string tension $\mu$, the energy density of long strings scales as $\rho_{\infty} \propto t^{-2}$. Since the background energy density also evolves as $\rho_{\rm bg} \propto t^{-2}$ for any equation-of-state parameter $\omega$, cosmic strings remain a fixed fraction of the total energy density and never dominate the Universe, unlike other cosmological defects.

During the fragmentation, long CS inter commute with themselves or with other strings and form loops with probability $P=1$. The loop formation rate is given as, 
\begin{align}
\frac{dn}{dt_{i}}= 0.1 \times C_{\rm eff}(t_{i})/(\alpha\, t^4_{i})\,,    
\end{align}
where $C_{\rm eff}$ is the efficiency factor and is derived within the framework of the Velocity-dependent One-Scale (VOS) model \cite{Martin:1996ea, Auclair:2019wcv}. Assuming an instantaneous scaling network, $C_{\rm eff}\sim 0.39$ for the matter-dominated era and $C_{\rm eff}\sim 5.4$ for the radiation domination. After its formation at time $t_{i}$ with length $l_{i}=\alpha t_{i}$, a loop shrinks through the emission of GWs \cite{Vachaspati:1984gt}, and its length evolve as 
\begin{eqnarray}
    l(t) = \alpha t_{i} - \Gamma G \mu (t-t_{i})\,.
\end{eqnarray}
Here, $\Gamma \simeq 50$ \cite{Vilenkin:1981bx} from Nambu-Goto simulations and $\alpha$ is the initial loop size parameter which takes a value $\alpha \sim 0.1$ preferred by numerical simulations \cite{Blanco-Pillado:2013qja, Blanco-Pillado:2017oxo}. The frequency of the emitted GWs at a time $\tilde{t}$ is related to length of the loops as 
\begin{eqnarray}
    \tilde{f} =\frac{2k}{l}\,,
\end{eqnarray}
where $k=1,2,3,......$ and denotes the different proper modes. The frequency today is related to the frequency at $\tilde{t}$ as $f=a(\tilde{t})/a(t_{0})\tilde{f}$. The energy emission rate of GWs from a loop in a particular mode $k$ depends on the small-scale structure of the loop and is given by, 
\begin{eqnarray}
    \frac{dE^{k}_{\rm GW}}{dt} = \Gamma^{k} G\mu^2\,,
\end{eqnarray}
where $\Gamma^{k}= \frac{k^{-\delta}\,\Gamma }{\sum_{p=1}^{p=\infty}p^{-\delta}}$ and $\sum_{k} \Gamma^{k}=\Gamma$. Here $\delta=4/3$ for small scale structure dominated by cusps and $\delta=5/3$ for kink domination~\cite{ Damour:2001bk}. In the following, we assume cusps-like structure to dominate the small-scale structure and only consider $\delta=4/3$. The resulting GWs energy density spectrum from the emission of all decaying loops is~\cite{Blanco-Pillado:2013qja},
\begin{eqnarray}
    \frac{d\rho_{\rm GW}(\tilde{f},\tilde{t})}{d\tilde{f}} = \int_{t_{\rm osc}}^{t_{0}} \frac{dE_{\rm GW}}{d\tilde{t}} \frac{dn(\tilde{f},\tilde{t})}{d\tilde{f}} d\tilde{t}\,,
\end{eqnarray}
with $t_{\rm osc}$ is the time when long strings start to oscillates and $\frac{dn(\tilde{f},\tilde{t})}{d\tilde{f}}$ is the loop number density per unit frequency and is given by 
\begin{eqnarray}
    \frac{dn(\tilde{f},\tilde{t})}{d\tilde{f}} = \left(\frac{a(t_{i})}{a(\tilde{t})}\right)^{3} \sum_{k} 0.1\times \frac{C_{\rm eff}(t_{i})}{(\alpha\, t^4_{i})} \frac{1}{\alpha (\alpha +\Gamma G \mu)} \times \frac{2k}{f^2} \left(\frac{a(\tilde{t})}{a(t_{0})}\right)^2\,.
\end{eqnarray}
With these, the final GWs energy density spectrum from CS can be expressed as \cite{Gouttenoire:2019kij}
\begin{eqnarray} \label{eq:CS_af}
    \Omega_{\rm GW} (f) &=& \frac{f}{\rho_{c}} \bigg|\frac{d\rho_{\rm GW}(f,t_{0})}{df}\bigg| \nonumber \\
    &\simeq& \frac{1}{\rho_{c}} \sum_{k} \frac{2k}{f} \frac{0.1\times\Gamma^{(k)}G\mu^2}{\alpha(\alpha + \Gamma G \mu)} \int_{t_{\rm osc}}^{t_{0}} \frac{C_{\rm eff}(t_{i})}{ t^4_{i}} \left(\frac{a(t_{i})}{a(\tilde{t})}\right)^3\left(\frac{a(\tilde{t})}{a(t_{0})}\right)^{5} \Theta(t_{i}-t_{\rm osc})\,d\tilde{t}\,.
\end{eqnarray}
Here, $\rho_{c}$ is the critical energy density today. The $\Theta$ denotes the Heaviside function and $t_{\rm osc}$ is the time at which long strings start oscillating, and we consider it to be the same as the time of CS network formation, $t_{\rm F}$ defined as $\sqrt{\rho_{\rm tot}(t_{\rm F})}\equiv\mu$. The contribution of the higher mode can be related to the contribution of the first mode by
\begin{eqnarray}
    \Omega^{(k)}_{\rm GW}(f) = \frac{\Omega^{(1)}_{\rm GW}(f/k)}{k^{4/3}}\,.
\end{eqnarray}

The formation and decay of loops during radiation domination yield a flat GW spectrum, whereas loops decaying during matter domination give a slope $f^{-1/3}$. In standard cosmology, the relation between observed frequency $f$ and the temperature $T$ of the thermal bath corresponding to the formation of loops can be found as 
\begin{eqnarray}\label{eq:f_T}
    f \simeq \sqrt{\frac{8}{z_{\rm eq}\,\alpha\, \Gamma\,G\mu}} \left(\frac{g_{*}(T)}{g_{*}(T_{0})}\right)^{1/4} \frac{T}{T_{0}}. \frac{1}{t_{0}},
\end{eqnarray}
where $z_{\rm eq}\simeq 3360$ is the redshift to the equality of matter and radiation of standard cosmology. The amplitude-frequency relation given by Eq.~\eqref{eq:CS_af} indicates the dependency of GWs spectrum on the expansion history of the universe. Hence, a non-standard cosmology can leave its imprint on the spectrum. In the presence of an early matter-dominated epoch, GWs spectrum starts to deviate from the standard cosmology at a frequency $f_{\Delta}$ called the turning point frequency~\cite{Cui:2018rwi, Blasi:2020wpy, Gouttenoire:2019kij, Gouttenoire:2019rtn} \footnote{Note that such spectral break can also appear when the energy loss of CS into particle production is more efficient than the GWs emission~\cite{Matsunami:2019fss, Auclair:2019jip}. Depending on the small scale structures (cusps or kinks), the breaks might appear in different places in the frequency spectrum. If one takes the approximate analytical formula given in ~\cite{Matsunami:2019fss} at face value, we find that  our scenario is better suited for kink-dominated loops which give less competitive contribution to the GWs spectra.}. Denoting $T_{\Delta}$ as the temperature when the most recent radiation era begins, the turning point frequency can be estimated as 
\begin{eqnarray}
    f_{\Delta} \simeq \sqrt{\frac{8}{z_{\rm eq} \alpha \Gamma G\mu}} \left(\frac{g_{*}(T_{\Delta})}{g_{*}(T_{0})}\right)^{1/4} \frac{T_{\Delta}}{T_{0}}. \frac{1}{t_{0}}.
\end{eqnarray}
For frequency $f> f_{\Delta}$, the GWs spectrum due to early matter domination fall as $\Omega_{\rm GW}(f) \propto (f/f_{\Delta})^{-1}$ for the fundamental $k=1$ mode and as $\Omega_{\rm GW}(f) \propto (f/f_{\Delta})^{-1/3}$ once higher modes are included.  
\begin{figure}
    \centering    \includegraphics[scale=0.84]{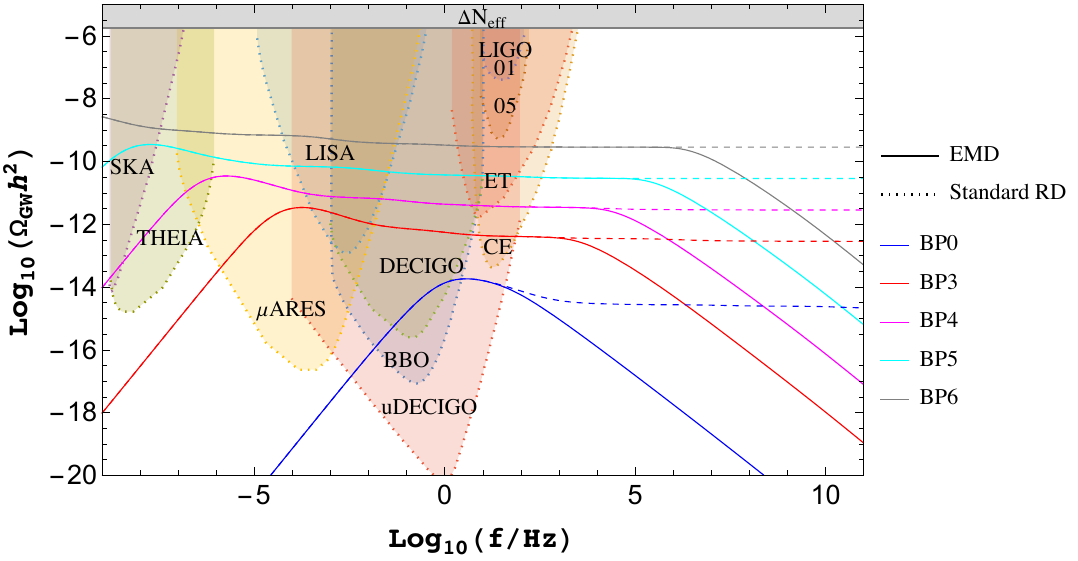}
    \caption{Spectrum of GW from CS for the BPs listed in Table~\ref{tab:BP}. The blue, red, magenta, cyan and gray lines represent the amplitude corresponding to BP0 ($G\mu \sim 1.5\times10^{-19})$, BP3 ($G\mu \sim 2\times10^{-15} $), BP4 ($G\mu \sim 2\times10^{-13} $), BP5 ($G\mu \sim 2\times10^{-11} $), and BP6 ($G\mu \sim 2\times10^{-9} $), respectively. The dashed lines distinguish the amplitude of the GW between the standard radiation domination and EMD. The color shaded regions denote current and future sensitivities of various experiments: LIGO~\cite{LIGOScientific:2014pky}, ET~\cite{Punturo_2010}, CE~\cite{LIGOScientific:2016wof}, DECIGO~\cite{Yagi:2011wg}, ultimate DECIGO ~\cite{Seto:2001qf}, BBO~\cite{Seto:2001qf}, LISA~\cite{LISACosmologyWorkingGroup:2022kbp}, $\mu$ARES~\cite{Sesana:2019vho}, THEIA~\cite{Garcia-Bellido:2021zgu} and SKA~\cite{Weltman:2018zrl}.}
    \label{fig:GW1}
\end{figure}

Fig.~\ref{fig:GW1} displays the GWs spectra corresponding to the benchmark points listed in Tab.~\ref{tab:BP} for the $k=1$ mode. The projected sensitivities of various current and future GW experiments are indicated by the shaded color regions. In the flat plateau region, the GW energy density scales as $\Omega_{\rm GW} \propto \sqrt{G\mu}$. Consequently, the amplitude decreases as one moves from BP6 to BP3, reflecting the reduction in the symmetry breaking scale. For BP1 and BP2, the corresponding value of $G\mu \sim 2.7\times10^{-35}$ and $G\mu \sim 10^{-26}$, respectively, leads to a GW spectrum that lies outside the frequency range of the future GW experiments considered here and is therefore not shown. Moreover, the turning point frequency from BP6 to BP3 decreases due to lower values of $T_{\Delta}$. We find that the turning-point frequencies for BP3 to BP6 listed in Tab.~\ref{tab:BP} lie outside the sensitivity ranges of future GW experiments, making it difficult to explicitly probe the effects of the EMD era through the spectral features alone. In principle, for a fixed value of $G\mu$, the turning-point frequency can be shifted into the sensitivity range of GW experiments by further lowering $T_{\Delta}$. However, we find that $T_{\Delta}$ cannot be lowered consistently while simultaneously maintaining the correct DM relic abundance. To illustrate how close the spectral break can be for a successful benchmark point, we also consider BP0, for which $T_{\Delta}$ is as low as $\sim 40$ MeV. In this case, the spectral break lies just within the projected sensitivity region of future uDECIGO experiments. Thus, the GW spectra from CS within the viable model parameter space provide a promising and potentially distinctive signature of the model. Finally, we also show the gray horizontal region excluded by CMB experiments due to the contribution of GWs to the extra effective number of relativistic species, denoted by $\Delta N_{\rm eff}$~\cite{Planck:2018jri,Caprini:2018mtu,Boyle:2007zx, Gouttenoire:2019kij}. Since the energy density of GWs redshifts in the same way as that of free radiation, inflationary PGWs can be treated as an additional radiation component and constrained through measurements of $\DNeff$. In terms of the present-day GW spectrum, the constraint can be written as~\cite{Maggiore:1999vm,Boyle:2007zx,Caprini:2018mtu}
\begin{align}\label{eq:GW-neff} & \left(\frac{h^2 \rho_\text{GW}}{\rho_c}\right)\Bigg|_0= \int_{f_\text{BBN}}^{\fmax}\,\frac{df}{f}\,h^2\,\ogw^{(0)}(f) \leq\frac{7}{8}\,\left(\frac{4}{11}\right)^\frac{4}{3}\,\Omega_\gamma^{(0)}\,h^2\DNeff\,,
\end{align}
which leads to $\ogw^{(0)}\,h^2\simeq 5.62\times 10^{-6}\,\DNeff$ where, $\Omega_\gamma^{(0)}\,h^2\simeq 2.47\times 10^{-5}$ is the relic photon abundance at the present epoch. Here, $f_{\rm BBN}$ denotes the present-day frequency of GWs produced at the time of BBN, which can be estimated using Eq.~\eqref{eq:f_T}. The maximum frequency, $f_{\rm max}$, corresponds to GWs emitted by the earliest population of cosmic-string loops, i.e., loops formed at the earliest time considered in the cosmic-string evolution.

Lastly, we define the signal-to-noise ratio (SNR), $\mathcal{S}$, by integrating over the experiment’s observation time $t_{\rm obs}$ and frequency window $[f_{\rm min}^{\rm exp},\,f_{\rm max}^{\rm exp}]$~\cite{Maggiore:1999vm,Allen:1996vm,Allen:1997ad}:
\begin{align}
\mathcal{S}=\sqrt{n_{\rm det}\,t_{\rm obs}\,\int_{f_{\rm min}^{\rm exp}}^{f_{\rm max}^{\rm exp}} df \left[\frac{\Omega_{\rm GW}(f)\,h^2}{\Omega_{\rm expt}(f)\,h^2}\right]^2}\,.
\end{align}
Here, $n_{\rm det}=1$ for auto-correlation searches and $n_{\rm det}=2$ for cross-correlation searches of the stochastic gravitational wave background (SGWB). A predicted GW signal is considered detectable if the resulting SNR exceeds the chosen threshold, which we take to be $\mathcal{S}>10$. 
\section{A collider-cosmology complementarity}
\label{sec:result}
\begin{figure}[htb!]
    \centering     \includegraphics[scale=0.72]{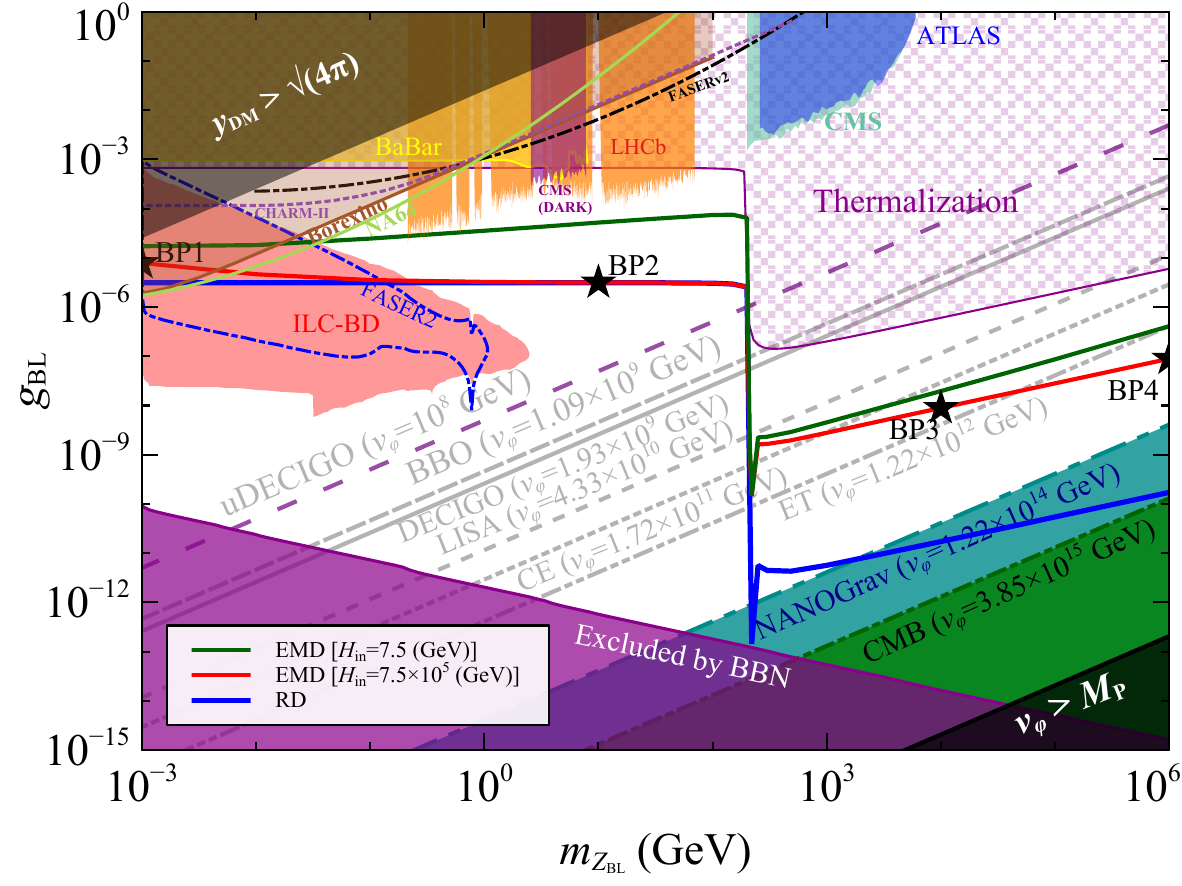}
    \caption{{\it Summary of parameter space:} contours of right DM abundance for $\mdm=100$ GeV, where we have fixed two values of the initial Hubble parameter: $\Hin=7.5\times10^5$ GeV (red solid line) and $\Hin=7.5$ GeV (green solid line). Along the blue shaded contour, the observed DM abundance is satisfied for production during RD. The diagonal gray contours represent SNR value $\mathcal{S}=10$, appearing from different GW experiments. All shaded regions are excluded from different experimental observations, as well as theoretical bounds (see text for details). BP1, BP2, BP3 and BP4, from Table \ref{tab:BP} are shown with black star points.}    
    \label{fig:relicparam}
\end{figure}

A summary of the viable parameter space, incorporating constraints from DM relic abundance, thermalization, BBN, and current/future collider and low-energy searches, is presented in Fig.~\ref{fig:relicparam}. Here we have compared DM production during $\zbl$-domination (EMD), considering $\Hin=\{7.5,\,7.5\times 10^5\}$ GeV shown via green and red curves respectively, with that during standard radiation domination, shown via the blue curve. The DM mass is fixed to $\mdm=100$ GeV, as a benchmark value. We also overlay the projected SNR contours for several next-generation GW experiments relevant to the modified GW signatures discussed earlier. 
\begin{figure}[htb!]
    \centering    \includegraphics[scale=0.72]{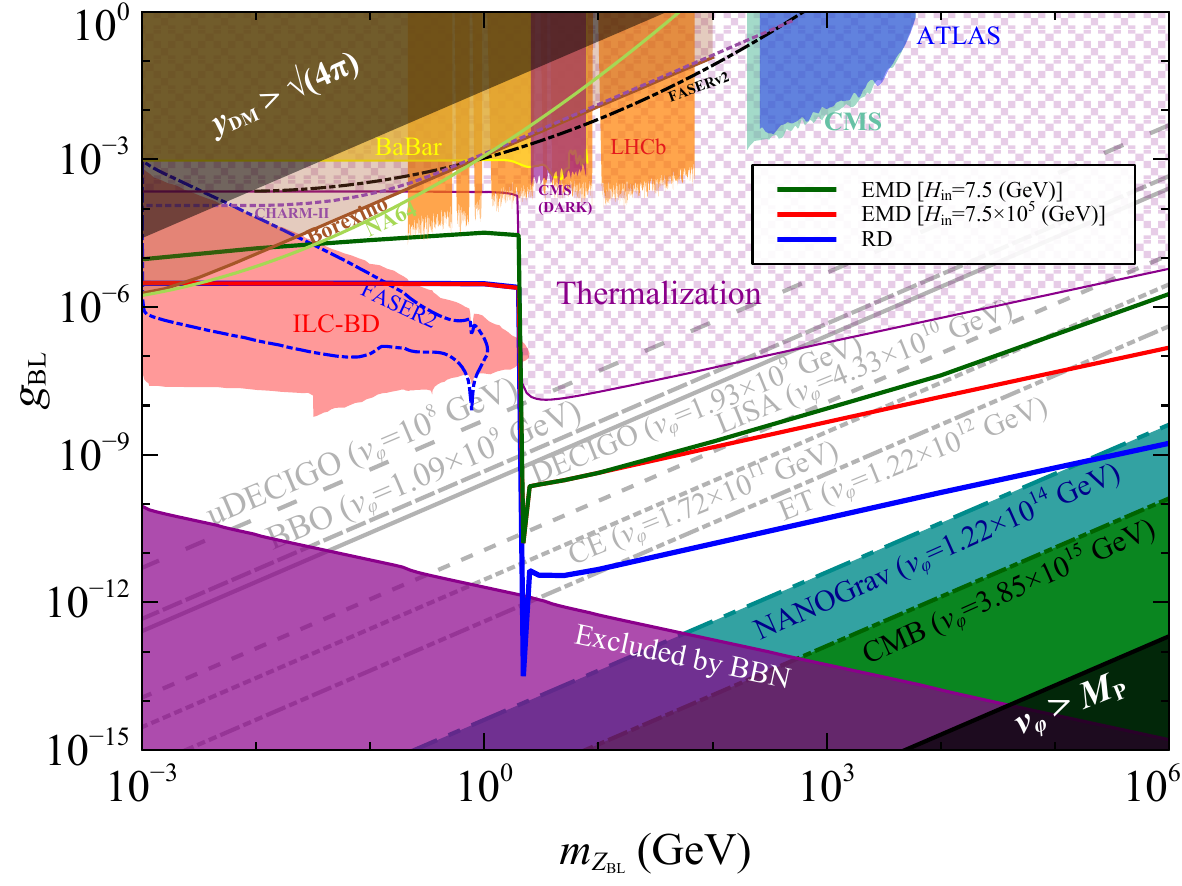}
    \caption{Same as Fig.~\ref{fig:relicparam}, but for $\mdm=1$ GeV.}  
    \label{fig:relicparam2}
\end{figure}

Let us first focus on the freeze-in relic contours for DM production during $\zbl$-domination, shown by the solid red curve. The freeze-in contour shows two characteristic regimes: (i) For $\Mzp < \mdm$, the relic abundance becomes largely insensitive to the mediator mass. In this region, DM production is dominated at temperatures $T \sim \mdm > \Mzp/2$, where the yield follows the third line of Eq.~\eqref{eq:Ydm}. Since this expression is independent of $\Mzp$, consistent with the last line of Eq.~\eqref{eq:gam-ann2}, the observed relic abundance primarily fixes the coupling $\gbl$, resulting in an almost flat contour in $\Mzp$. (ii) For $\Mzp > 2\mdm$, DM production is dominated by on-shell $\zbl$ decay. Here, the asymptotic yield scales as $Y_0 \propto \gbl^2/\Mzp^{16/3}$ (for fixed EMD parameters), as seen from Eq.~\eqref{eq:Ndm12}. Therefore, larger mediator masses require stronger couplings to reproduce the observed DM abundance. Using Eq.~\eqref{eq:gbl}, we identify the magenta-shaded region with $\mathbb{R}>1$ (labeled ``Thermalization"), where the DM thermalizes during the EMD epoch and hence is excluded. For comparison, the blue solid curve shows the parameter region that reproduces the observed relic abundance when DM is produced during the standard radiation-dominated era. In this case, the final DM yield from $\zbl$-mediated scattering is given by,
\begin{align}\label{eq:y0RD}
Y_0^{\rm RD}\propto\gbl^4
\begin{dcases}
\frac{M_P\,\left(\Trh^{\rm RD}\right)^3}{\Mzp^4}\,, & T\ll\Mzp/2\,,
\\[10pt]
\frac{M_P}{\Gamma_{\zbl}}\,, & T\simeq\Mzp/2\,,
\\[10pt]
\frac{M_P}{T_0}\,, & T\gg\Mzp/2\,,
\end{dcases}
\end{align}
where $\Trh^{\rm RD}$ denotes the maximum temperature of the Universe at the onset of radiation domination. In obtaining Eq.~\eqref{eq:y0RD} we have analytically solved Eq.~\eqref{eq:beq} using the standard RD Hubble rate, $\mathcal{H}_{\rm RD}=\left(\pi/3\right)\sqrt{\gs(T)/10}\,(T^2/M_P)$, considering DM production via $\zbl$-mediated 2-to-2 scattering of the bath particles. For a very heavy mediator, the DM yield depends on the highest temperature reached during the RD epoch, exhibiting the characteristic UV freeze-in behavior. In contrast, when the mediator is sufficiently light, corresponding to the regime $\Mzp<2\,\mdm$, the yield becomes effectively independent of the mediator mass, as observed previously. For $\Mzp>2\,\mdm$ (decay-dominated production), the coupling required to reproduce the observed relic density is smaller than in the EMD scenario. This is a consequence of the slower expansion rate during radiation domination compared to EMD, due to which a larger coupling is required for freeze-in production during EMD, compared to standard cosmology.

A distinctive aspect of the present scenario is that the EMD epoch is not introduced phenomenologically, but arises from the dynamics of the B$-$L gauge boson. As a result, the properties of the EMD phase are directly correlated with the experimentally constrained parameter space $[\gbl-\Mzp]$, allowing current observations to probe non-standard cosmological evolution before BBN. A noteworthy consequence of DM production during an EMD era is that reproducing the observed relic abundance necessitates a larger DM–SM coupling, simply because the Hubble rate is larger compared to standard radiation-domination. This enhances the observational prospects of the underlying model and naturally leads to a complementarity between collider experiments and cosmological searches. 

For $\Mzp \lesssim 1$ GeV, we find that the parameter space consistent with the observed DM relic abundance for production during the $\zbl$-dominated epoch is already excluded by existing constraints from LHCb~\cite{Foguel:2022unm,Baruch:2022esd}, BaBar~\cite{BaBar:2014zli,BaBar:2017tiz}, and CMS~\cite{CMS:2023slr,Asai:2023mzl,CMS:2024zqs} for $\gbl \gtrsim 10^{-4}$, corresponding to lower values of $\Hin\lesssim 7.5$ GeV. Although the region associated with larger $\Hin$ remains beyond the reach of current searches from LHCb, BaBar, CMS and ATLAS, but it falls within the projected sensitivity of the proposed ILC beam-dump (ILC-BD)~\cite{PhysRevD.103.035024,Asai:2023dzs} and FASER~\cite{FASER:2018eoc} experiments, which are expected to probe couplings as small as $\gbl \sim 10^{-8}$. Furthermore, existing limits from BoreXino~\cite{Borexino:2018ohr,Borexino:2020umd}, CHARM-II~\cite{CHARMII:1993,Vilain:1994qy} and Na64~\cite{Banerjee:2019hmi,Banerjee:2017hhz}, together with the projected sensitivities of FASER-2~\cite{Feng:2018pew,Anchordoqui:2022fpf}, exclude $\gbl \gtrsim 10^{-6}$ for $\Mzp \lesssim 0.1$ GeV, covering both the low- and high-$\Hin$ regions. Overall, collider and beam-dump experiments either already exclude or will be able to probe the parameter space corresponding to DM production during the EMD epoch for sub-GeV $\Mzp$, with sensitivity reaching couplings as small as $\gbl \sim 10^{-8}$. In contrast, future GW observatories such as BBO~\cite{Crowder:2005nr,Corbin:2005ny}, DECIGO~\cite{Seto:2001qf,Kudoh:2005as}, LISA~\cite{LISA:2017pwj}, Cosmic Explorer (CE)~\cite{Reitze:2019iox}, and Einstein Telescope (ET)~\cite{Hild:2010id,Punturo:2010zz,Sathyaprakash:2012jk,Maggiore:2019uih} are more efficient in constraining EMD scenario corresponding to very feeble coupling, typically, $\gbl\lesssim 10^{-9}$, depending on $\Mzp$. Meanwhile, existing constraints on cosmic string tension from NANOGrav~\cite{NANOGrav:2023hvm} already exclude the decay-dominated DM production scenario during standard radiation domination for $\Mzp\gtrsim 200$ GeV with $\gbl\sim 10^{-12}$. Similarly, existing constraints on cosmic string tension from CMB observations~\cite{Charnock:2016nzm,Lizarraga:2016onn} excludes even smaller $\gbl\lesssim 10^{-13}$ for $\Mzp\sim 1$ TeV, with the bound getting stronger as $\Mzp$ increases. Since smaller $\gbl$ results in long-lived $\zbl$, that can spoil the standard BBN prediction, hence we impose $\tau_{\zbl}\equiv\Gamma_{\zbl}^{-1}>1$ sec, to ensure the $\zbl$ decay is complete before the onset of BBN. This excludes the bottom left part of the parameter space (labeled as ``Excluded by BBN"), depending on the choice of $[\Mzp,\,\gbl]$ that results in $\tau_{\zbl}>1$ sec. Further, for $\Mzp\gtrsim 10$ TeV and $\gbl\lesssim 10^{-15}$, the VEV $v_\phi$ becomes super-Planckian [cf. Eq.~\eqref{eq:mass}]. This excludes the bottom right panel of the parameter space (labeled as $v_\phi>M_P$). From Eq.~\eqref{eq:mass}, requiring the RHN Yukawa coupling to remain perturbative, $Y_N < \sqrt{4\pi}$, leads to an upper bound on the B$-$L breaking scale, $v_\phi \equiv v_\phi^{\rm pert} < \mdm/\sqrt{8\,\pi}$. Consequently, the upper-left region of the parameter space (shaded gray) is excluded, as it corresponds to $\Mzp/(2\,\gbl) > v_\phi^{\rm pert}$.

The IceCube neutrino observatory has reported the detection of three neutrino events with energies in the PeV range, representing an excess of roughly $3\sigma$ above the expected atmospheric background~\cite{IceCube:2013cdw, IceCube:2013low, IceCube:2014stg,IceCube:2015gsk,IceCube:2015qii,Li:2025tqf}. These events correspond to deposited energies of 1.04~PeV, 1.14~PeV, and 2.0~PeV. Although the origin of such ultra high energy neutrinos remains uncertain, it has been suggested that they could arise from the decay of superheavy DM particles~\cite{Bai:2013nga,Higaki:2014dwa,Esmaili:2014rma,Murase:2015gea,Dudas:2018npp,Cohen:2016uyg,Rott:2014kfa}. In scenarios where DM undergoes a two-body decay involving a neutrino, the resulting neutrino spectrum exhibits a characteristic high-energy cutoff at approximately half the DM mass~\cite{Higaki:2014dwa,Esmaili:2014rma}. Furthermore, observations from IceCube impose a robust lower bound on the DM lifetime, $\tau_{\rm DM}\gtrsim \mathcal{O}(10^{28})\,\mathrm{s}$~\cite{Esmaili:2014rma,Arguelles:2022nbl}. This constraint is largely model-independent and implies a lifetime many orders of magnitude greater than the age of the Universe.
\begin{figure}[htb!]
    \centering    \includegraphics[scale=0.72]{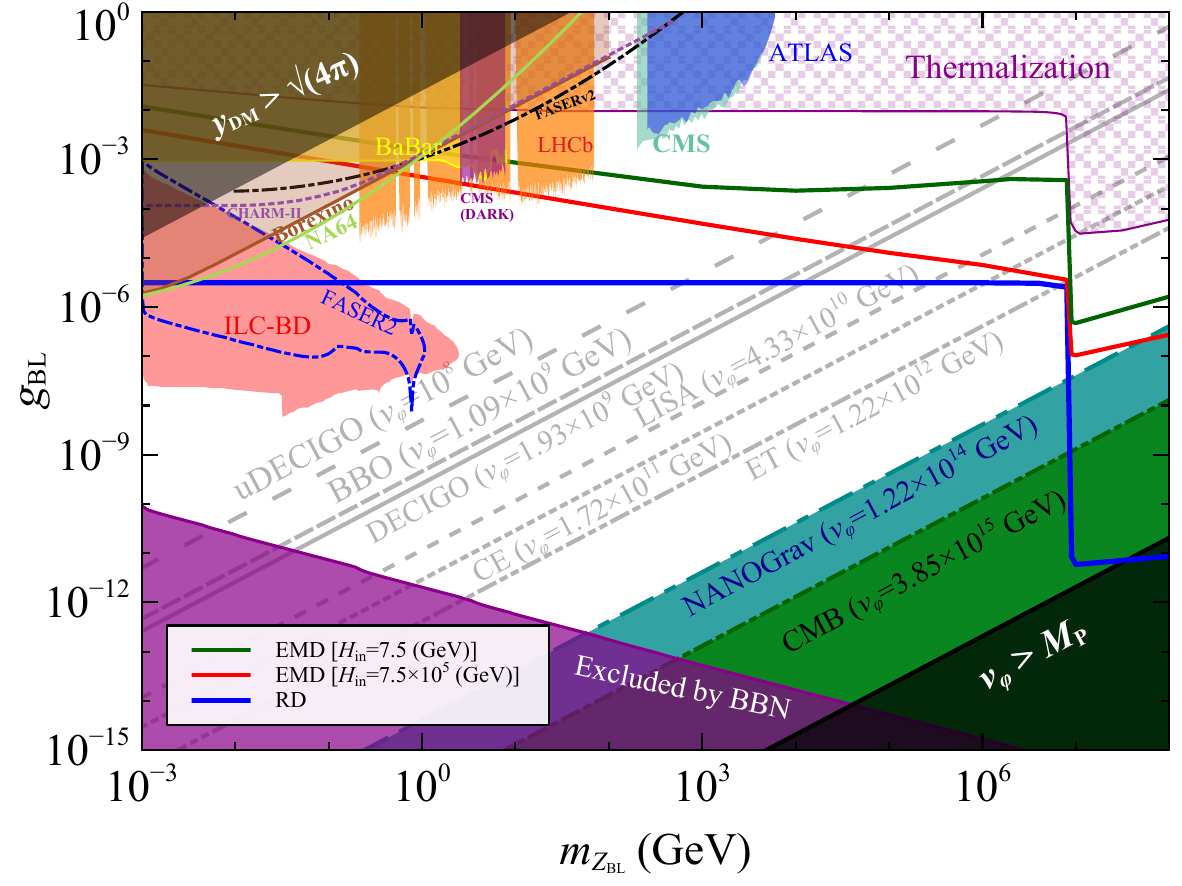}
    \caption{Same as Fig.~\ref{fig:relicparam}, but for $\mdm=4$ PeV to explain the IceCube neutrino events with a decaying DM.}  
    \label{fig:icecube}
\end{figure}

In the present framework, a decaying RHN DM candidate with mass 4 PeV can be produced during the EMD epoch. The corresponding parameter space yielding the observed relic abundance is shown in Fig.~\ref{fig:icecube}. Interestingly, depending on the value of $\Hin$, a portion of the viable parameter space (dominated by $\zbl$-mediated scattering) is excluded by existing constraints from BaBar, LHCb, CMS and perturbative bound, particularly for $\gbl \gtrsim \mathcal{O}(10^{-3})$ and $\Mzp \lesssim \mathcal{O}(100)$ GeV. If the IceCube neutrino excess is indeed sourced by the decay of PeV-scale DM produced during an EMD epoch via $\zbl$-mediated scattering, the resulting constraints become progressively weaker for larger values of $\Hin$. In fact, for sufficiently large $\Hin$, the relic density prediction converges to that of the standard radiation-dominated scenario (shown in blue), reflecting the diminishing impact of the EMD phase. This can be understood from the fact that increasing $\Hin$ results in higher $\Tmax$ [cf. Eq.~\eqref{eq:Tmax}], resulting efficient DM production. Hence to satisfy the right abundance, a smaller $\gbl$ is required (for the same $\Mzp$) such that the EMD lasts for a longer epoch causing dilution of the excess relic. Thus, the coupling required for the observed abundance approaches the standard RD result. We note that the relic abundance contour corresponding to DM production during the reheating era exhibits a negative slope, particularly for smaller values of $\Mzp$. This behavior can be understood as follows. For the hierarchy $\mdm \gg \Mzp$, the DM production is dominated by temperatures close to the maximum bath temperature, $\Tmax$. Since $\Tmax$ increases with $\Mzp$, as evident from Eq.~\eqref{eq:Tmax}, a larger mediator mass enhances the DM production rate. Consequently, the observed relic abundance can be reproduced with a smaller value of the gauge coupling $\gbl$, giving rise to the negative slope of the relic contour. Once again, we notice, future GW observations can constrain a large range of the relic density allowed parameter space, typically for $\Mzp\gtrsim 10$ TeV and $\gbl\lesssim 10^{-5}$, considering scattering-dominated DM production. Finally, for $\Mzp>2\,\mdm=8$ PeV, production via on-shell $\zbl$ decay opens up, making corresponding $\gbl\lesssim 10^{-7}$, to satisfy the observed abundance. Such feeble coupling can still be constrained from existing bounds due to NANOGrav and CMB, for $\Mzp\gtrsim 10^8$ GeV, depending on the choice of $\Hin$. 

Because the DM particle is Majorana in nature, it couples axially to the $\zbl$ gauge boson, whereas quarks interact with $\zbl$ through vector couplings. Consequently, the $\zbl$-mediated contribution to DM--nucleon scattering is velocity-suppressed and vanishes in the non-relativistic limit relevant for direct-detection experiments. The dominant contribution therefore arises from spin-independent elastic scattering mediated by the Higgs bosons. The corresponding DM--proton cross section is given by,
\begin{align}
\sigma_{\rm SI}^p
=\frac{4}{\pi}
\left(\frac{m_p\,\mdm}{m_p+\mdm}\right)^2
f_p^2\,,
\end{align}
where the effective DM--proton coupling is
\begin{align}
\frac{f_p}{m_p}
=
\sum_{q=u,d,s}
f_{T_q}^p\,\frac{\alpha_q}{m_q}
+\frac{2}{27}
\left(1-\sum_{q=u,d,s}f_{T_q}^p\right)
\sum_{q=c,b,t}
\frac{\alpha_q}{m_q}\,.
\end{align}
The effective DM--quark interaction strength is
\begin{align}
\alpha_q
=
-\frac{\gbl\,y_q}{2\sqrt{2}}
\,\frac{\mdm}{\Mzp}\,
\sin 2\theta
\left(
\frac{1}{m_1^2}
-\frac{1}{m_2^2}
\right),
\end{align}
where $m_q$ and $y_q$ denote the quark mass and Yukawa coupling, respectively, while $m_{1(2)}$ correspond to the masses of the SM-like (singlet-like) Higgs bosons. Since our analysis is performed in the limit of vanishing scalar mixing ($\theta \to 0$), the effective coupling $\alpha_q$ becomes highly suppressed. As a result, the Higgs-mediated DM--nucleon interaction is effectively negligible, rendering the spin-independent scattering cross section far below the sensitivity of current direct-detection experiments. 

Let us now briefly comment on the generation of the observed baryon asymmetry in this framework. So far, we have focused only on the production of the DM candidate, $N_1$, during the $\zbl$-dominated era. The two heavier right-handed neutrinos, $N_2$ and $N_3$, can be efficiently produced through the $N_{2,3}LH$ Yukawa interactions when $T \gtrsim m_{N_{2,3}}$, where $m_{N_{i}}$ denotes the mass of $N_i$. Their production from on-shell $\zbl$ decays is kinematically forbidden because $\Mzp < 2m_{N_{2,3}}$. Assuming a hierarchical mass spectrum with $m_{N_{2}} \ll m_{N_{3}}$, the out-of-equilibrium, \textit{CP}-violating decays of $N_2$ can generate the required lepton asymmetry~\cite{Fukugita:1986hr}. Under these assumptions, the observed baryon asymmetry can be explained through standard leptogenesis without affecting the DM production mechanism or the allowed DM parameter space. Therefore, we do not attempt a quantitative analysis of baryogenesis in the present work\footnote{Leptogenesis in non-standard cosmological histories has been studied extensively in the literature (see, for example, Refs.~\cite{Abdallah:2012nm,Dutta:2018zkg,Mahanta:2019sfo,Chen:2019etb,DiMarco:2022doy}).}. This framework can thus simultaneously accommodate successful DM production and baryogenesis during the EMD epoch.
\section{Conclusions}
\label{sec:concl}
In this work, we study an anomaly-free gauged B$-$L extension of the Standard Model (SM). Gauging the B$-$L symmetry introduces a heavy Abelian gauge boson whose energy density is assumed to dominate the Universe before Big Bang Nucleosynthesis. This leads to an early matter-dominated (EMD) era, modifying the expansion history of the Universe relative to the standard radiation-dominated cosmology. We identify the lightest right-handed neutrino (RHN) as a viable decaying dark matter (DM) candidate and show that its observed relic abundance can be produced through the freeze-in mechanism during the EMD epoch. The heavier RHNs, $N_{2,3}$, simultaneously generate light neutrino masses via the type-I seesaw mechanism, while their CP-violating out-of-equilibrium decays produce the observed baryon asymmetry through leptogenesis. An important feature of DM production during an EMD era is that achieving the correct relic abundance requires a larger DM--SM coupling than in the standard radiation-dominated Universe. This enhances the prospects of probing the model at collider and beam-dump experiments, with the accessible parameter space depending on the mass of the new gauge boson and the strength of its gauge coupling. The scenario can also be tested through primordial gravitational waves generated by cosmic strings, whose future observation could significantly constrain the model. Furthermore, if the heavy DM produced during EMD decays into ultra-high-energy neutrinos, experiments such as IceCube could provide complementary information on both the underlying particle physics model and the early cosmological history. Overall, our study demonstrates how gravitational wave observations, neutrino telescopes, collider searches, and cosmological measurements together provide a powerful multi-messenger approach to exploring physics beyond the Standard Model.
\section*{Acknowledgment}
BB and ND would like to acknowledge the hospitality at IIT, Hyderabad during the WHEPP-2025 workshop, when this work was initiated. BB would like to acknowledge SRM University-AP for support through the Internal Research Grant Policy. P. K. P. acknowledges the Ministry of Education, Government of India, for providing financial support for his research via the Prime Minister’s Research Fellowship (PMRF) scheme. The authors would like to thank Arindam Das for providing the collider constraints and Rome Samanta for useful discussions and suggestions regarding GWs.
\appendix
\section{Relevant cross-sections \& decay rates}
\label{sec:decay-cross-section}
Following are the $\zbl$-mediated SM,\,SM $\to$ DM,\,DM cross-sections
\begin{align}    &\sigma_{\ell\ell\rightarrow N_1N_1}=\frac{g_{\rm BL}^4}{24\pi s}\frac{(s-4m_{N_1}^2)(s+2m_\ell^2)}{(s-4m_\ell^2)\left[(s-m_{Z_{\rm BL}}^2)^2+\Gamma_{Z_{\rm BL}}^2m_{Z_{\rm BL}}^2\right]}\sqrt{(s-4m_\ell^2)(s-4m_{N_1}^2)}\,,
\\&
\sigma_{qq\rightarrow N_1N_1}=\frac{g_{\rm BL}^4}{648\pi s}\frac{(s-4m_{N_1}^2)(s+2m_q^2)}{(s-4m_q^2)\left[(s-m_{Z_{\rm BL}}^2)^2+\Gamma_{Z_{\rm BL}}^2m_{Z_{\rm BL}}^2\right]}\sqrt{(s-4m_q^2)(s-4m_{N_1}^2)}\,,
\end{align}
where the $\zbl$ decay rates are given by,
\begin{align}
&\Gamma_{Z_{\rm BL}\rightarrow \ell\bar{\ell}}=
\frac{g_{\rm BL}^2}{12\pi}\,
m_{Z_{\rm BL}}
\left(1+2\frac{m_\ell^2}{m_{Z_{\rm BL}}^2}\right)
\left(1-\frac{4m_\ell^2}{m_{Z_{\rm BL}}^2}\right)^{1/2}\,,
\\&
\Gamma_{Z_{\rm BL}\rightarrow q\bar{q}}=
\frac{g_{\rm BL}^2}{36\pi}\,
m_{Z_{\rm BL}}
\left(1+2\frac{m_q^2}{m_{Z_{\rm BL}}^2}\right)
\left(1-\frac{4m_q^2}{m_{Z_{\rm BL}}^2}\right)^{1/2}\,,
\\&
\Gamma_{Z_{\rm BL}\rightarrow N_iN_i}=
\frac{g_{\rm BL}^2}{24\pi}\,
m_{Z_{\rm BL}}
\left(1-\frac{4m_{N_i}^2}{m_{Z_{\rm BL}}^2}\right)^{3/2}\,.
\end{align}
Here, $\ell,\,q$ are the SM leptons and quarks, respectively.
\bibliography{Bibliography}
\bibliographystyle{JHEP}
\end{document}